\documentclass{aa}
\usepackage{longtable,graphicx,amssymb,natbib,rotating,psfrag,amsmath,epsfig,array,txfonts,subfigure}

\begin{document}

\title{High energy emission and polarisation limits for the \textit{INTEGRAL}\thanks{Based on
    observations with \textit{INTEGRAL}, an ESA project with instruments and
    science data centre funded by ESA member states (especially the PI
    countries: Denmark, France, Germany, Italy,
    Switzerland, Spain), Czech Republic and Poland, and with the participation
    of Russia and the USA.} burst GRB\,061122}

\author{S. McGlynn\inst{1,2}
 \and
S. Foley\inst{1} \and
B. McBreen\inst{1} \and
L. Hanlon\inst{1} \and
S. McBreen\inst{1,3} \and
D.~J. Clark\inst{4} \and
A.~J. Dean\inst{4} \and
A. Martin-Carrillo\inst{1}\and
R. O'Connor\inst{1} }

\offprints{S. McGlynn, \email{smcglynn@particle.kth.se}}

\institute{UCD School of Physics, University College
  Dublin, Dublin 4, Ireland
  \and Department of Physics, Royal Institute of Technology (KTH), AlbaNova University Centre, SE-10691 Stockholm, Sweden
\and Max-Planck-Institut f\"{u}r extraterrestrische Physik, D-85741 Garching, Germany
\and School of Physics and Astronomy, University of Southampton, Southampton,
SO17 1BJ, UK}

\date{Received 5 September 2008 / Accepted 24 March 2009}
\abstract
{GRB\,061122 is one of the brightest GRBs detected within \textit{INTEGRAL}'s field of view to date, with a peak flux (20--200 keV) of 32 photons cm$^{-2}$ s$^{-1}$ and fluence of 2 $\times$ 10$^{-5}$ erg cm$^{-2}$. Polarisation measurements of the prompt emission are relatively rare. The spectral and polarisation results can be combined to provide vital information about the circumburst region.}
{The $\gamma$--ray detectors on \textit{INTEGRAL} were used to investigate the spectral characteristics of GRB\,061122. A search for linear polarisation in the prompt emission was carried out. The X-ray properties were examined using data from the X--Ray Telescope (XRT) on \textit{Swift}.}
{The $\gamma$--ray properties of GRB\,061122 were determined using IBIS and SPI. The multiple event data of GRB\,061122 from SPI were analysed and compared with Monte--Carlo simulations using the \textit{INTEGRAL} mass model. The $\chi^2$ distributions between the real and simulated data as a function of the percentage polarisation and polarisation angle were calculated.}
{The prompt spectrum was best fit by a combination of a blackbody and a power--law model, with evidence for high energy emission continuing above 8 MeV. A pseudo-redshift value of \textit{pz} = 0.95 $\pm$ 0.18 was determined using the spectral fit parameters.  The jet opening angle was estimated to be smaller than $2.8^\circ$ or larger than $11.9^\circ$ from the X-ray lightcurve. An upper limit of 60\% polarisation was determined for the prompt emission of GRB\,061122.}
{The high energy emission observed in the spectrum may be due to the reverse shock interacting with the GRB ejecta when it is decelerated by the circumburst medium. This behaviour has been observed in a small fraction of GRBs to date, but is expected to be more commonly observed by the Fermi Gamma-ray Space Telescope. The conditions for polarisation are met if the jet opening angle is less than $2.8^\circ$, but further constraints on the level of polarisation are not possible.}

\titlerunning{Properties of GRB\,061122}
\authorrunning{McGlynn et al.}
\maketitle

\section{Introduction\label{intro}}
Long $\gamma$--ray bursts (GRBs) are linked to the collapse of a massive star which forms a rapidly rotating
black hole \citep{piran04,mesz2006}. In addition, a large ordered magnetic field may be induced by the angular
momentum of the accretion disk \citep{Zhang2004}. Energetic outflows develop, which are beamed
perpendicular to the accretion disk and along the black hole's rotation
axis. A GRB is detected if the observer is close to the jet axis. Polarisation
is generally associated with an asymmetry in the way that the material is
viewed. The asymmetry can be attributed to a preferential orientation of the
magnetic field, the geometry of the source or the surrounding environment \citep{lazz2006}. The link between the $\gamma$--ray production
mechanism and the degree of linear polarisation can be exploited to constrain models of GRB production.

Most bright GRB spectra can be fit by the Band model \citep{band:1993} which is an empirical function comprising two smoothly broken power--laws, with the distributions of the low energy and high energy power--law photon indices around values of $\alpha=-1$ and $\beta=-2.2$ respectively \citep{yuki06}. A thermal component of the prompt emission has also been proposed \citep[e.g.][]{ghirlanda:2003,ryde:2004,ryde05}. This model is a hybrid of the Planck black body function plus a simple power--law model and is of the form:

\begin{equation}
N(E)= A \left(\frac{E^{2}}{{\rm exp\left(\frac{E}{kT}\right)-1}}\right) + B E^{\alpha}
\end{equation}

\noindent where $kT$ represents the black body temperature in keV, and $\alpha$ represents the power--law index.

The thermal emission may originate from the transition from opaque to transparent in a wind photosphere \citep{lyut00,dm02}. Most GRB spectra are dominated by non--thermal radiation corresponding to the synchrotron/inverse Compton emission generated in the optically thin environment, usually interpreted as the signature of internal shocks. The relative strengths of the thermal and non--thermal components can vary with time over the burst duration. In some cases, the thermal (i.e. black--body) component is dominant in the first few seconds of the burst \citep{ryde:2004} and decreases in strength so that the power--law component dominates the later emission.

\textit{INTEGRAL} \citep{wink2003} has observed 52 long--duration GRBs ($T_{90}\gtrsim2$~s, \cite[e.g.][]{sf07}) and one short GRB ($T_{90}\lesssim2$~s, \citet{me08}) to the end of June 2008. The spectral and temporal properties of the most intense burst detected, GRB\,041219a, have been previously published \citep{mcbreen06}. The level of polarisation was also determined for GRB\,041219a using multiple event data from the spectrometer (SPI) on board \textit{INTEGRAL} \citep{kal07,me06}. SPI was not specifically designed as a polarimeter, but polarisation can be measured through observed multiple scatter events due to the layout and geometry of the detector array. \textit{RHESSI} is the only other instrument currently in orbit with the ability to measure $\gamma$--ray polarisation \citep{wigger04}. 

In this paper we present the results of the $\gamma$--ray spectral and temporal characteristics of the intense burst GRB\,061122 obtained with SPI and the Imager (IBIS) onboard \textit{INTEGRAL} (\S~\ref{spec_res}). The results of polarisation analysis using the SPI multiple event data of GRB\,061122 are presented in \S~\ref{pol_an}, using the method described in \cite{me06}. We also present afterglow results from \textit{Swift}-XRT (\S~\ref{ag}). The implications of the spectral analysis and limit on the polarisation are discussed in \S~\ref{disc}.

The cosmological parameters adopted throughout the paper are $H_0$ = 70\,km\,s$^{-1}$\,Mpc$^{-1}$, $\Omega_m = 0.3$, $\Omega_{vac} = 0.7$. We adopt the notation for the $\gamma$--ray spectra that $\alpha$ represents the low energy power--law photon index and the power--law index in the quasithermal model, $\beta$ represents the high energy power--law photon index and $E_{peak}$ is the peak energy of the spectral fit. The power--law photon index of the X--ray spectrum is represented by $\Gamma_X$ and the temporal slope is given by $\alpha_X$. All errors are quoted at the 1$\sigma$ confidence level.

\section{\textit{INTEGRAL}\label{int}}
The Spectrometer on \textit{INTEGRAL} (SPI) consists of 19 hexagonal germanium (Ge) detectors covering the energy range 20~keV--8~MeV. The fully coded field of view (FoV) is 16$^{\circ}$ corner--to--corner, with a partially coded FoV of 34$^{\circ}$. A detailed description of SPI is available in \cite{ved2003}. The event data from SPI are separated into single events where a photon deposits
energy in a single detector, and multiple events where the photon Compton scatters and deposits
energy in two or more detectors. The single events are used for spectral and temporal analysis, while the multiple events are used for polarisation analysis. The failure of detectors 2 and 17 reduces the effective area to about 90\% of the original area for single events. It is reduced to $\sim$ 75\% for multiple events, because the number of adjacent detector pairs drops from 84 to 64.

The imager IBIS consists of two separate detector layers, ISGRI (energy range 15~keV--1~MeV) and PICsIT (energy range $\sim$
180~keV--10~MeV). A detailed description of IBIS can be found in \cite{uber2003}. The ISGRI detector is made up of 16384 CdTe pixels, creating a pixellated imager with good spatial resolution and decreased spectral resolution compared with SPI (8~keV at 100~keV). The fully coded field of view is 9$^{\circ}\,\times\,9^{\circ}$, with a coded mask 3.4~m above the detector plane.  The \textit{INTEGRAL} Burst Alert System (IBAS, \citet{ibas}) detects and localises $\sim$ 1 GRB/month utilising data from the ISGRI detector.

The two $\gamma$--ray instruments on \textit{INTEGRAL} are suitable for spectral analysis. Data from SPI and IBIS were used to determine the spectral characteristics of GRB\,061122, while multiple event data from SPI were used in the polarisation analysis.

\section{Prompt and Afterglow Observations\label{obs}}

\begin{table}[t]
\caption{Properties of GRB\,061122 obtained with \textit{INTEGRAL}.}
\label{table:grbs}
\centering
\begin{minipage}{0.9\textwidth}
\begin{tabular}{cccc}
\hline\hline
\it R.A.\footnote{Coordinates are taken from the relevant GCN circular\\ \citep{gcn5834}.} & \it Dec. & \it Off--Axis & \it Trigger   \\
              &        & \it Angle & \it Time (UTC)   \\ 
\hline
 & & & \\
 20h 15m 20.9s & +15$^{\circ}$ 30' 50.8'' & 8.2$^{\circ}$ & 07:56:45 \\
 & & & \\
& & & \\
\hline\hline
 $T_{90}$ & \multicolumn{2}{c}{\it Peak Flux (20--200~keV)}  &\\ 
\it (s) &  \it (ph\,cm$^{-2}$\,s$^{-1}$) & \it (erg\,cm$^{-2}$\,s$^{-1}$)  & \\
\hline
& & & \\
 11 &31.69 $^{+0.65}_{-0.93}$ & 3.13 $^{+0.06}_{-0.09} \times 10^{-6}$  & \\
 & & & \\
\hline
\end{tabular}
\end{minipage}
\end{table}

GRB\,061122 was detected by IBAS at 07:56:45 on 22 November 2006, at a location
of R.A. =  20h 15m 20.9s, Dec = +15$^{\circ}$ 30' 50.8'' \citep{gcn5834}. GRB\,061122 was a bright burst with an initial fluence reported in the 20--200~keV range of $3\,\times\,10^{-6}$ erg cm$^{-2}$ \citep{gcn5836} and a peak flux of 31.7 ph cm$^{-2}$ s$^{-1}$, making it the second most intense burst observed by \textit{INTEGRAL} after GRB\,041219a. KONUS--Wind also triggered on the burst, and reported a fluence of $2.31 ^{+0.05}_{-0.12}\,\times\,10^{-5}$ erg cm$^{-2}$ in the energy range 20~keV--2~MeV \citep{gcn5841}. 

The GRB location was observed by XRT on \textit{Swift} starting approximately 7 hours post--trigger \citep{gcn5846,rep_17} where a fading X--ray afterglow with a flux of $\sim 3\,\times\,10^{-12}$ erg cm$^{-2}$ s$^{-1}$ was observed. Using 2245~s of overlapping XRT Photon Counting mode and UVOT V--band data, the astrometrically corrected X--ray position was R.A. = 20h 15m 19.79s, Dec.  = +15$^{\circ}$ 31' 02.3" with an uncertainty of 2.0", consistent with the \textit{INTEGRAL} location. 
R--band observations of the error region of GRB\,061122 were taken on two consecutive nights using the
MDM 2.4m telescope in Arizona \citep{gcn5849}. A fading object was discovered within 1" of the X--ray afterglow candidate. The observations are listed in Table~\ref{opt}. The magnitudes were not corrected for Galactic extinction which is estimated to be A$_{R} = 0.49$ mag.

\begin{table}[t]
\caption{R--band observations of the optical afterglow of GRB\,061122 from the MDM telescope.}
\label{opt}
\centering
\begin{tabular}{lccc}
\hline\hline
\it Date (UT)    &    \it  Time (UT)   & \it T - T$_0$ (s)   & \it R$_{mag}$ \\
\hline
 Nov. 23   &   01:52   &    17.9  &   22.61 $\pm$ 0.05\\
 Nov. 24   &   02:26   &    42.5  &   23.41 $\pm$ 0.15\\
\hline
\end{tabular}
\end{table}

\section{Gamma--ray Spectral and Temporal Analysis\label{sa}}

\subsection{Lightcurves}
\begin{figure}[t]
\begin{center}
\includegraphics[width=0.7\columnwidth,angle=270]{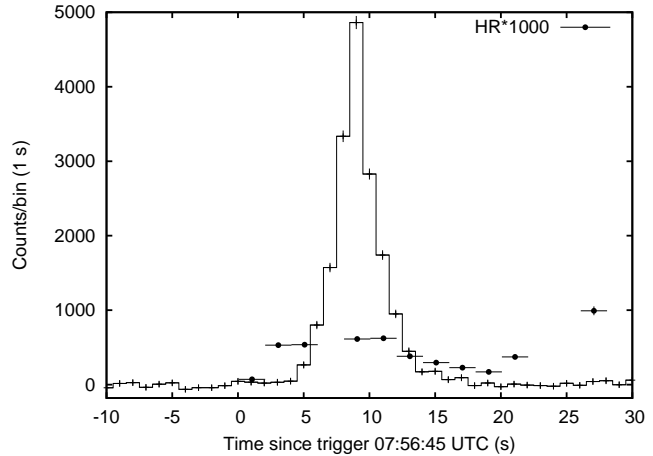}
\caption{Background--subtracted SPI lightcurve of GRB\,061122 in the energy range 20~keV--8~MeV at 1~s resolution. The hardness ratios between the energy ranges 25--100\,keV and 100--300\,keV calculated from IBIS data are overlaid (circles). The hardness ratios are multiplied by 1000 for clarity of presentation.
\label{lcs}}
\end{center}
\end{figure}

The background--subtracted SPI lightcurve of GRB\,061122 is presented in Fig.~\ref{lcs} and the lightcurves per SPI detector are shown in Fig.~\ref{spi_dets1122}. All lightcurves are in 1 second bins with the trigger time, T$_0$, at 07:56:45. GRB\,061122 is composed of a single relatively symmetric pulse. The KONUS Wind lightcurve\footnote{http://www.ioffe.rssi.ru/LEA/GRBs/GRB061122\_T28608/} also shows a single pulse of approximately the same duration as \textit{INTEGRAL}. There are significant telemetry gaps in the IBIS data (T$_0$+1 -- T$_0$+5, T$_0$+6 -- T$_0$+9), so a higher resolution lightcurve could not be generated. SPI was not affected by these telemetry gaps. The burst was observed in all of the SPI detectors (Fig.~\ref{spi_dets1122}), making it a possible candidate for polarisation analysis. The hardness ratio between 25--100\,keV and 100--300\,keV was calculated for each 2\,s interval of the burst using IBIS data because SPI does not have sufficient energy resolution, and shows initial hard to soft evolution followed by hardening after the main emission episode (Fig.~\ref{lcs}). 

\subsection{Spectral Analysis\label{SA}}
The spectra were extracted using specific GRB tools from the
Online Software Analysis \citep{diehl2003,skin2003} version 5.1 
available from the \textit{INTEGRAL Science Data Centre}. The T$_{90}$
duration (the time for 5\%--95\% of the GRB counts to be recorded) was
determined using the lightcurve generated from the IBIS/ISGRI data in 1~s bins. The
T$_{90}$ interval was then selected for the spectral analysis in both instruments. The SPI data was fit over the energy range 20\,keV -- 8\,MeV and the IBIS data from 20\,keV -- 1\,MeV. Table~\ref{table:grbs} lists the details of GRB\,061122, including the off--axis angle, T$_{90}$, and peak flux obtained with SPI in the 20--200\,keV energy range.
 
Each spectrum was fit with several spectral models: a simple power--law (PL), the Band model (GRBM, \citet{band:1993}), a combination of a blackbody and simple power--law model (BB+PL, \citep[e.g.][]{ryde05}) and a cutoff power--law which is a variation of the Band model with $\beta = \infty$ (Cutoff PL). The spectra from IBIS and SPI were also fit simultaneously (Joint Fit), with the normalisation between the two instruments free to vary. The parameters and fluences from each fit are listed in Table~\ref{spec1122}. The burst was divided into 2 second intervals and spectral analysis was performed with SPI. These results are listed in Table~\ref{rt}. KONUS--Wind \citep{konus} also triggered on GRB\,061122 and the spectral results are listed in Table~\ref{spec1122} for comparison.

\section{Spectral Results\label{spec_res}}
\begin{figure}[t]
\begin{center}
\includegraphics[width=0.65\columnwidth,angle=270]{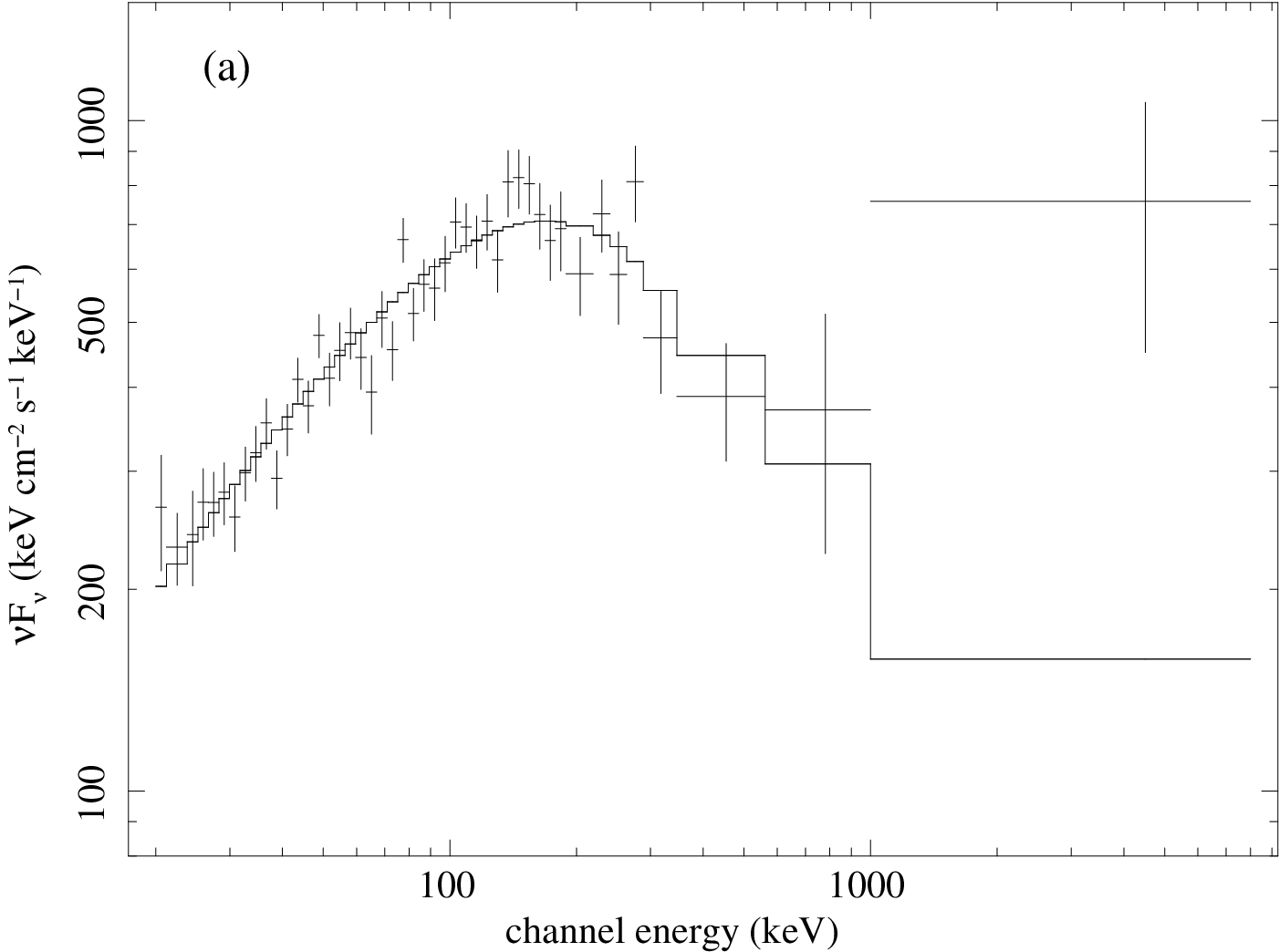}
\includegraphics[width=0.65\columnwidth,angle=270]{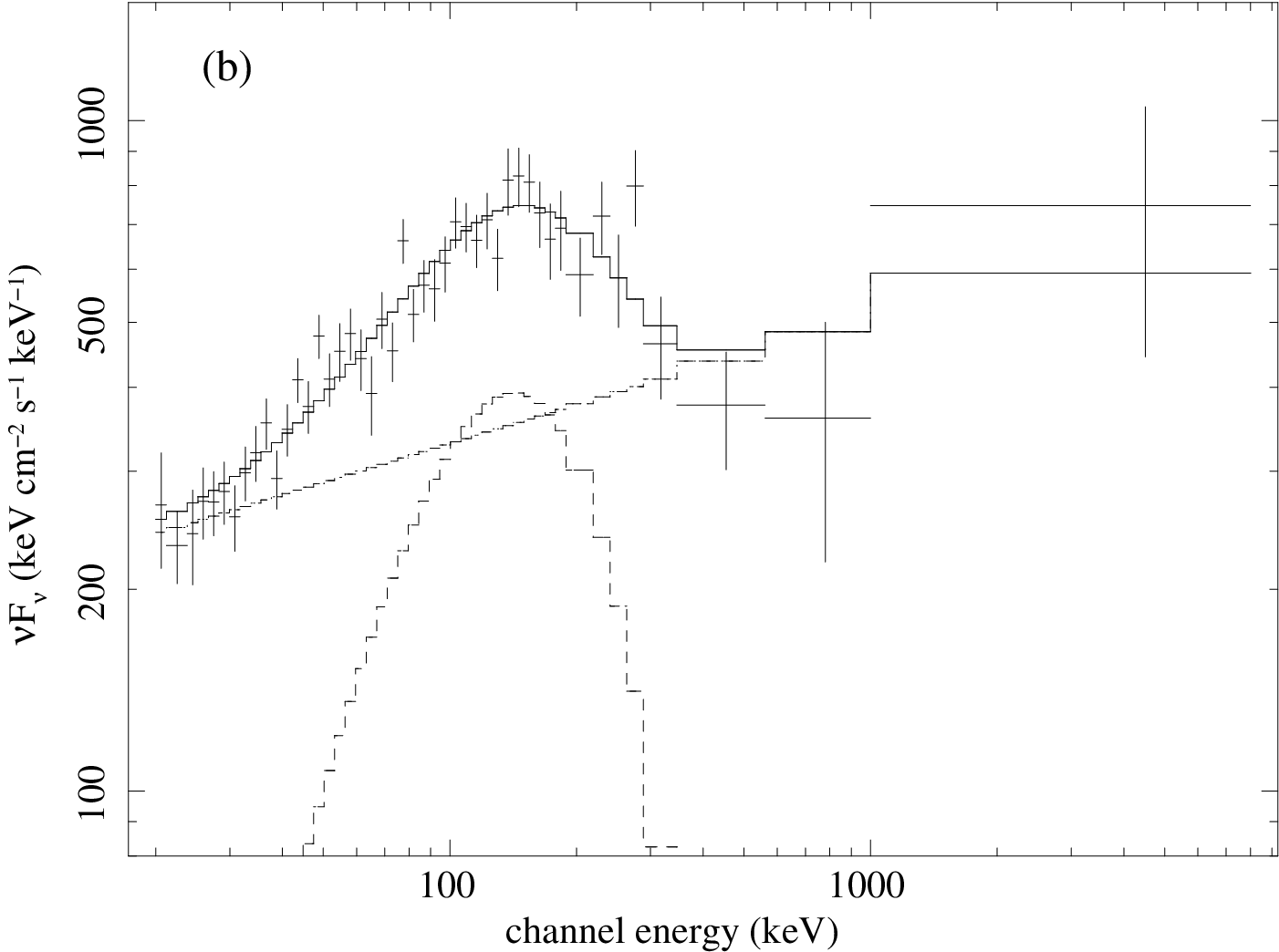}
\caption{The $\nu$F$_{\nu}$ SPI spectra over a 12 second interval of  GRB\,061122 (a) fit by the Band model, indicating an excess of counts at the high energy range and (b) fit by a blackbody and power--law model. The data have been rebinned for clarity of presentation.\label{spectra:figs}}
\end{center}
\end{figure} 

\begin{table*}[t]
\centering
\caption{Spectral fit parameters for GRB\,061122 with spectral models as described in \S\ref{SA} and reduced $\chi^2$ per degrees of freedom (dof). \label{spec1122}}
\begin{minipage}{\textwidth}
\centering
\begin{tabular}{c|l|cccccc}
\hline\hline
\it Detector &\it Spectral  & \it $\alpha$\footnote{low energy photon index in the Band model/single power--law index for the PL/BB+PL/Cutoff PL models} & \it $\beta$ & \it E$_{0}$/kT  &  \it E$_{peak}$ &\it $\chi^2_{red}$ & \it Fluence\footnote{20--200~keV}  \\
&\it Model & & & \it (keV) & \it (keV) & \it /dof & \it ($ 10^{-6}$ erg cm$^{-2}$) \\
\hline
& & & & & & & \\
& PL & -1.67 & ... & ... & ... & 3.92/58 & 17.79 $^{+0.23}_{-0.32}$ \\
SPI\footnote{20\,keV -- 8\,MeV} & GRBM & $-0.98^{+0.11}_{-0.12}$ & $-2.72^{+0.34}_{-0.85}$ &
166$^{+39}_{-28}$ & 169$^{+44}_{-35}$& 1.19/56 & 19.63 $^{+0.16}_{-0.71}$ \\
& BB+PL & $-1.81^{+0.08}_{-0.07}$ & ...  & 36$\pm 3$ & ... &1.21/56 & 19.96 $^{+0.54}_{-0.71}$\\
& Cutoff PL & $-1.01^{+0.10}_{-0.11}$ & ... & ... & 179$^{+38}_{-28}$ & 1.24/57 & 19.64 $^{+0.64}_{-1.47}$\\
& & & & & & & \\
& PL & -1.47 & ... & ... & ... & 2.30/92 & 5.13 $^{+0.09}_{-0.11}$ \\
Joint Fit & GRBM & $-1.14^{+0.27}_{-0.32}$ & $-1.91^{+0.07}_{-0.10}$  & 81$^{+120}_{-70}$ & 70$^{+106}_{-63}$ & 1.35/90 & 8.80 $^{+0.11}_{-4.62}$ \\
(IBIS \& SPI) & BB+PL & $-1.74 \pm 0.07$ & ... & 15$^{+4}_{-3}$ & ... & 1.40/90 & 8.36 $^{+0.04}_{-0.29}$ \\
& Cutoff PL & $-0.97^{+0.12}_{-0.13}$ & ... & ... & 129$^{+45}_{-29}$ & 1.45/91 & 5.10 $^{+0.33}_{-0.60}$\\
& & & & & & & \\
KONUS\footnote{KONUS--WIND spectral parameters are taken from \cite{gcn5841} and are in the energy range 20\,keV--2\,MeV.} & Cutoff PL & $-1.03^{+0.06}_{-0.07}$ & ... & ... & 160$^{+8}_{-7}$ & 1.03/62 & 23.1 $^{+0.5}_{-1.2} $ \\
& & & & & & & \\
\end{tabular}
\end{minipage}
\end{table*}

The spectra were fit with the models described in \S~\ref{sa}. The fit parameters for each model are listed in Table~\ref{spec1122}. The values of $\alpha$, the low energy photon index, and $\beta$, the high energy photon index, are consistent with the distribution of values obtained by \citet{yuki06}. The simple power--law model (PL) is not as good a fit as the models with curvature, since a break is visible in the spectrum (Fig.~\ref{spectra:figs}). There is also evidence for a high
energy excess, which is better fit by the blackbody + power--law model (BB+PL, Fig.~\ref{spectra:figs}~(b)). The IBIS/SPI joint fits were not as good as the SPI spectrum on its own, since the SPI spectrum was finely binned and much better fits were obtained than with IBIS. The spectral results for IBIS are not included in the table because the gaps in the data interfered with the fitting. The same effect rendered the joint fit poorer than that of the SPI data. The reduced $\chi^2$ is close to 1 for the SPI spectral fits and although the GRBM has a better reduced $\chi^2$, the BB+PL model seems to better account for the high energy emission.

The high energy component persists for up to 5 seconds after the burst. The fluence from 15--20\,s after the trigger is $\sim 8 \times 10^{-7}$ erg\,cm$^{-2}$ in the 1--8\,MeV energy range compared to $\sim 4.4 \times 10^{-7}$ erg\,cm$^{-2}$ in the 20--200\,keV energy range.

\begin{table}[h]
\centering
\caption{SPI spectral parameters of GRB\,061122 in 2 second intervals during the burst, fit by the Band model and combined blackbody and power--law model.}
\label{rt}
\begin{minipage}{\textwidth}
\begin{tabular}{llrrll}
\hline\hline
\it Time  & \it Model  & $\alpha$    &  $\beta$   & $E_{peak}$ & \it Fluence\footnote{20--200\,keV}\\
\it from T$_0$ & &  & &/\emph{kT} &\it  $\times\,10^{-6}$\\
\it (s) & & & &\it (keV) & \it (erg cm$^{-2}$) \\
\hline
 5--7 & PL & $-1.81 ^{+0.16}_{-0.15}$ & ... & ... & $1.41$ \\
 7--9 & GRBM & $-0.69 ^{+0.20}_{-0.18}$ & $-2.56 ^{+0.36}_{-0.42}$ & $200 ^{+82}_{-71}$& $7.61$  \\
 & BB+PL & $-1.66 ^{+0.11}_{-0.09}$ & ... & $41 \pm 4$ & $7.81$ \\
 9--11 & GRBM & $-0.52 ^{+0.15}_{-0.20}$ & $-2.93 ^{+0.29}_{-2.05}$ & $138 ^{+53}_{-57}$& $8.56$ \\
 & BB+PL & $-1.79 ^{+0.15}_{-0.20}$ & ... & $31\pm 3$ & $8.67$ \\
11--13 & GRBM & $-1.23 ^{+0.36}_{-0.24}$ & $-3.0$\footnote{fixed} & $114 ^{+80}_{-45}$ & $2.78$ \\
 & BB+PL & $-2.10 ^{+0.36}_{-0.24}$ & ... & $31 \pm 8$ & $2.83$ \\
& & & & & \\
\end{tabular}
\end{minipage}
\end{table}

\citet{vian08} have recently published the IBIS spectral results of GRB\,061122 and also note the presence of the data gaps. They obtained the best fit to the IBIS data with a cutoff power--law with parameters $\alpha = -1.24 \pm 0.16$ and $E_0 = 122 ^{+60}_{-31}$\,keV. These values are consistent with the results from SPI and from the joint fit presented in Table~\ref{spec1122}.

The burst was divided into 2 second intervals and the spectral analysis was carried out for each interval using SPI data. The fit results are listed in Table~\ref{rt}. The peak energy in the Band model fit decreases with time and $\beta$ steepens. The value of \emph{kT} decreases from 41\,keV to 31\,keV and the photon index of the BB+PL fit evolves from $-1.66$ to $-2.10$ through the burst. However, the overall values in each fit are mainly consistent within the error bars.

The KONUS--WIND spectrum was also fit by a cutoff power--law model over the brightest 12 seconds \citep{gcn5841} and the spectral fits obtained (in the 20~keV--2~MeV range) are also listed in Table~\ref{spec1122}. The peak flux on a 64--ms time scale measured over 3 seconds from KONUS was 8.81 $^{+0.83}_{-1.05}$$\,\times\,10^{-6}$ erg cm$^{-2}$ s$^{-1}$. The fit parameters from KONUS are in good agreement with the cutoff power--law fit from SPI, with the KONUS fit in the energy range 20\,keV--2\,MeV and the SPI fit in the range 20\,keV--8\,MeV. 

\section{Polarisation\label{pol_an}}

\subsection{Model Simulations for Polarisation in SPI\label{model}}
\begin{figure}[t]
\begin{center}
\includegraphics[width=0.6\columnwidth,angle=90]{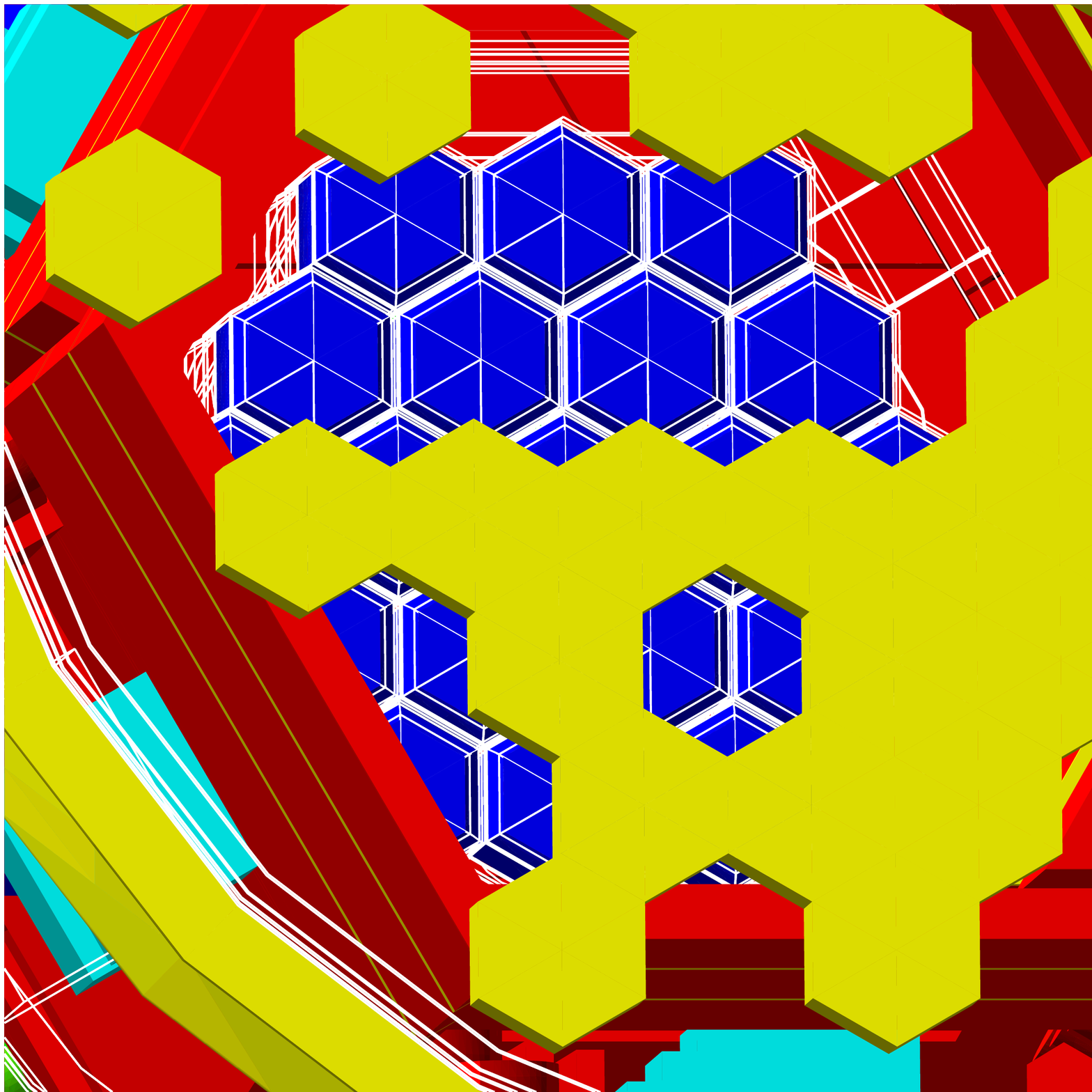}
\caption{The coded mask elements (yellow) overlaying the 19 SPI detectors (blue), as viewed from the direction of the incoming GRB photons generated using the simulations. Detectors 14, 15 and 16 (bottom left) are partially obscured by the anticoincidence shield. 
\label{sim}}
\end{center}
\end{figure}

The dominant mode of interaction for photons in the energy range of a few hundred keV is Compton scattering.  Linearly polarised $\gamma$--rays preferentially scatter
perpendicular to the incident polarisation vector, resulting in an azimuthal
scatter angle distribution which is modulated relative to the distribution for
unpolarised photons. The 19 segmented detectors in SPI (Fig.\,\ref{sim}) register the scattering of events into multiple detectors. Using a combination of real data collected from SPI and simulated data, it is possible to calculate the level of polarisation present in a GRB.

A computer model of the \textit{INTEGRAL} spacecraft written in the GEANT 4
toolkit \citep{Agostinelli} was used to simulate SPI multiple events. This model was developed from the GEANT
3 \textit{INTEGRAL} Mass--Model (TIMM) \citep{2003A&A...411L..19F} originally designed to provide background and performance evaluation of all the instruments onboard
\textit{INTEGRAL}. The model contains an accurate representation of the SPI
instrument, including the mask and veto elements. The rest of the spacecraft
is modelled to a much lower level of detail. 

\begin{figure*}[t]
\begin{center}
\includegraphics[width=0.85\columnwidth,angle=270]{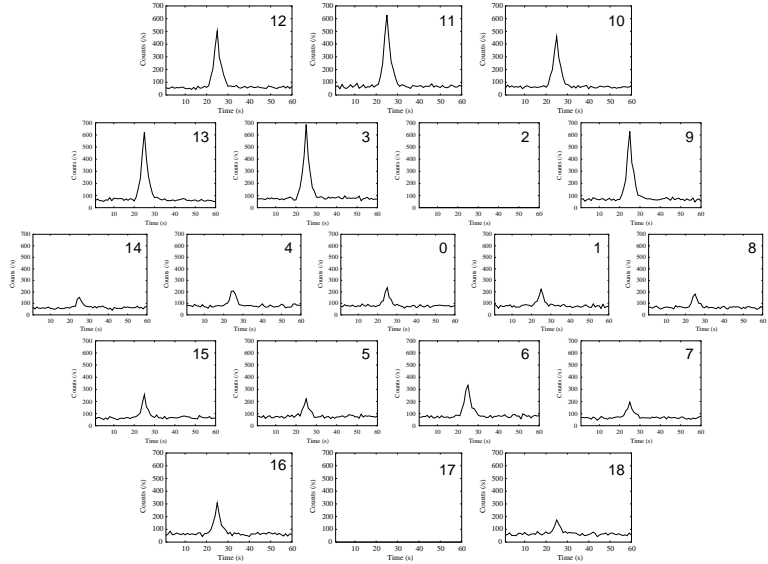}
\caption{The layout of the 19 detectors of SPI with single event lightcurves of GRB\,061122
  showing the variation in count rate per detector. GRB\,061122 is evident in all operational detectors and the weakest signals
are in detectors partially or fully covered by the mask. Detectors 14, 15 and 16 are also partially obscured by the anticoincidence shield.
\label{spi_dets1122}}
\end{center}
\end{figure*}

The simulation of the GRB multiple events was carried out as described in \cite{me06}. The spectral parameters from the Band model of the $T_{90}$ spectrum were used to generate a set of simulated events arriving from the direction of the GRB. For each simulation run, the polarisation angle of the photons was
set between 0$^{\circ}$ and 180$^{\circ}$ in 10 degree
steps, and the polarisation fraction was set to 100\%. There was one run for a beam of unpolarised photons. The unpolarised
simulation data were combined with the polarised simulation data, allowing the
percentage of polarisation to be varied as a function of angle. 

The polarisation analysis procedure for GRB\,061122 was carried out in the same manner as for GRB\,041219a in \cite{me06}. The SPI multiple event data were divided into six directions in the energy ranges of 100--350~keV and 100--500~keV using the kinematics of the Compton scatter interactions, and divided into 3 directions in the 100--1000~keV energy range. The number of multiple events between 100--350\,keV, 100--500\,keV and 100\,keV--1\,MeV were 244, 303 and 927 respectively for GRB\,061122. The total number of simulated events was $\sim 10^5$ per energy range.
  
These event lists were compared with the simulated data from the \textit{INTEGRAL} mass model and the value of $\chi^2$ was calculated for a
range of polarisation angles and percentages of polarisation. These values were used to generate
significance level contour plots, which gave a minimum for the angle and
percentage of polarisation that most closely matches the actual data. The
results of the fitting procedure are given in Table~\ref{table_res}, which
lists the best fit percentage polarisation and the angle for the GRB in the energy ranges
100--350~keV, 100--500~keV and 100~keV--1~MeV. The errors quoted for the
percentage and angle of polarisation are 1$\sigma$
for 2 parameters of interest. 

\begin{table*}[t]
\centering
\caption{Table of results from $\chi^2$ fitting of real and simulated
  data. \label{table_res}}
\begin{minipage}{\textwidth}
\centering
\begin{tabular}{l|cc||ccc}
\hline\hline
Polarisation\footnote{Errors quoted are 1$\sigma$
for 2 parameters of interest. The columns from left to right list the polarisation percentage, angle and best--fit probability that
  the model simulations matched up with the real data, the energy ranges analysed over six directions (columns 2 and 3) and the energy ranges analysed over three directions (columns 4--6).} & \multicolumn{2}{c||}{6 Directions} & \multicolumn{3}{c}{3 Directions} \\
 & 100--350~keV & 100--500~keV & 100--350~keV & 100--500~keV & 100~keV--1~MeV\\ 
\hline
 & & & & & \\
 Percentage (\%) & $>$ 31 & $>$ 32 & 11 $ ^{+48}_{-11}$ & 25 $^{+45}_{-25}$  & 29 $^{+25}_{-26}$   \\
 Angle & $>$ 40 & $>$ 90 & $>$ 40 & 100 $ ^{+65}_{-66}$ & 100 $ ^{+32}_{-24}$ \\
 Probability (\%) & 26.4 & 28.0 & 99.8 & 99.4 & 97.3 \\
 & & & & & \\
 \hline
\end{tabular}
\end{minipage}
\end{table*}

\begin{figure}[t]
\begin{center}
\includegraphics[width=0.75\columnwidth]{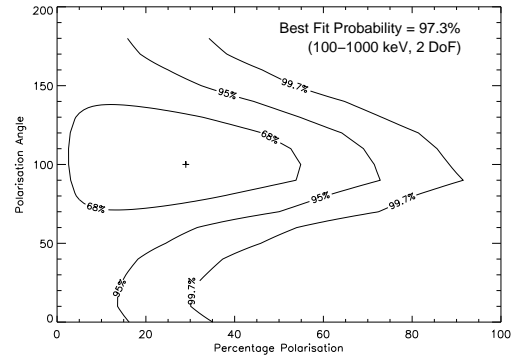}
\caption{Contour plot of the percentage
  polarisation as a function of the
polarisation angle for the three scatter directions ($\rm{0}^{\circ}-\rm{180}^{\circ}$) of GRB\,061122, showing the 68\%, 95\% and
  99.7\% probability contours in the energy ranges 100--1000~keV.
\label{cont}}
\end{center}
\end{figure}

GRB\,061122 occurred at 8$^{\circ}$ off-axis and the detector plane was almost completely illuminated (Fig.~\ref{spi_dets1122}) with the largest count rates observed in the detectors at the top of the plane (detectors 10--12). The six direction data provide poorer polarisation constraints due to the low statistics. The background scatter is also non-linear, which contributes to the smearing-out of the polarisation signal. The best fit probability that
  the model simulations provide a good description of the real data is $\sim$ 97\% for the three scatter directions in all 3 energy ranges (Table~\ref{table_res}), corresponding to an upper percentage polarisation limit of 60\%. The contour plot for the 100--1000\,keV energy range is shown in Fig.~\ref{cont}. Only the 1\,$\sigma$ contour is closed indicating the paucity of statistics available.

\section{Afterglow Analysis\label{ag}}
\begin{figure}[t]
\begin{center}
\includegraphics[width=0.7\columnwidth]{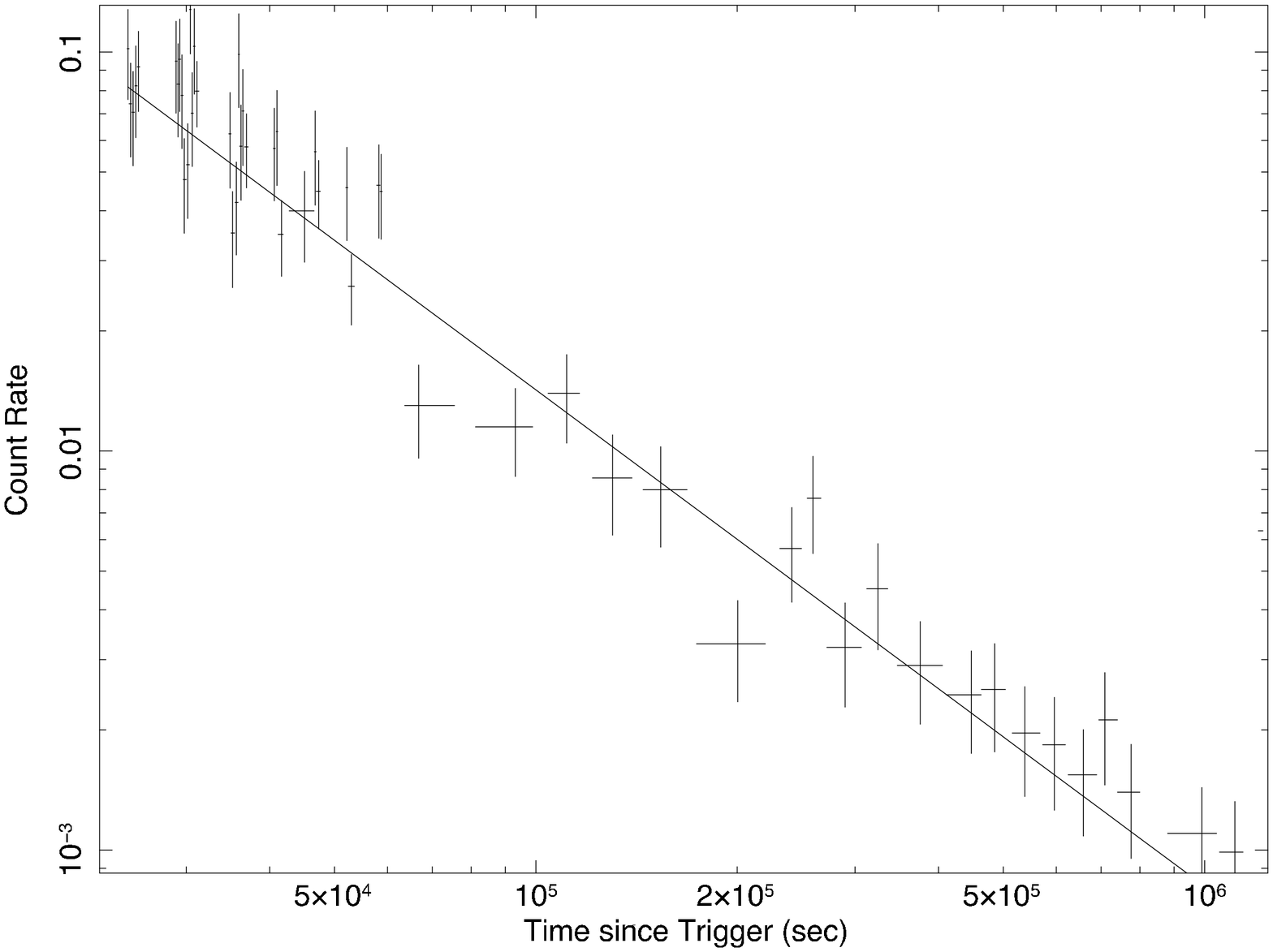}
\includegraphics[width=0.6\columnwidth,angle=270]{0920_6b.ps}
\caption{Top panel: XRT lightcurve of GRB\,061122 fit with a power--law slope of $-$1.24 $\pm$ 0.07. Lower panels: XRT count rates and hardness ratio in the energy ranges 0.3--1.5 and 1.5--10\,keV.
\label{Swift_lc_061122}}
\end{center}
\end{figure}

The 0.3--10~keV X-ray lightcurve  was fit with a decaying power--law with a slope of $\alpha_X = -1.24 \pm 0.07$ over the time interval T$_0$ + 24.5~ks to T$_0$ + 76~ks (Fig.~\ref{Swift_lc_061122}). The presence of another nearby source contaminated the XRT lightcurve at late times, so the source extraction region was reduced to minimise contamination.

The X--ray spectrum over the interval  T$_0$ + 24.5~ks to T$_0$ + 1267~ks was fit by an absorbed power--law with a photon index of $\Gamma_X = -2.02 \pm 0.16$ and a column density of 2.15 $\pm$ 0.45 $\times 10^{21}$ cm$^{-2}$, comparable to the Galactic column density in the direction of the source (1.5\,$\times\,10^{21}$ cm$^{-2}$). The average unabsorbed 0.3--10~keV flux for this spectrum is 2.5 $^{+0.3}_{-0.5}$ $\times 10^{-13}$ erg cm$^{-2}$ s$^{-1}$. 

The XRT hardness ratio is shown in Fig~\ref{Swift_lc_061122}. There appears to be significant spectral hardening from about 10$^5$~s to the end of the observation. However,  when the spectra were subdivided into early and late times, the spectral parameters could not be significantly constrained, due to contamination.

\section{Discussion\label{disc}}

\subsection{Constraints on Redshift and Luminosity\label{rl}}
An estimate of the redshift, the `pseudo-redshift', can be obtained using the burst spectral parameters. \citet{pel08} used a sample of HETE-II GRBs with known $z$ to test the pseudo-redshift calculation and found that the dispersion of the ratios between the spectroscopically measured redshift and the pseudo-redshift was smaller than a factor of 2. The pseudo--redshift was calculated for GRB\,061122 using the SPI Band model fits from Table~\ref{spec1122} and the online pseudo--redshift calculator and was found to be $pz = 0.95 \pm 0.18$. 

Using this pseudo--redshift and the spectral fluence and peak flux from \S \ref{spec_res}, the isotropic peak luminosity L$_{iso}$ was estimated to be $1.47 \pm 0.05 \times 10^{52}$ ergs s$^{-1}$ (50--300\,keV) and the isotropic equivalent bolometric energy E$_{iso}$ = $8.5 \pm 3.6 \times$ 10$^{52}$ erg (1--1000\,keV). 

\subsection{The Hard Tail Component of GRB\,061122}
GRB\,061122 exhibits a high energy spectral component throughout the duration of the burst (Fig.~\ref{spectra:figs}). The high energy component does not turn over within the energy range of SPI, indicating that emission may exist above $\sim$\,8\,MeV. In addition, this hard component may persist up to 5 or more seconds after the main emission pulse. A significant fluence ($\sim 8 \times 10^{-7}$ erg\,cm$^{-2}$, 1--8\,MeV) is present in the 5 seconds after the emission appears to end at T$_0 + 20$s in Fig\,\ref{lcs}.

GRB\,941017 was the first burst with a significant long lasting high energy component detected up to $\sim$\,200\,MeV, discovered by EGRET \citep{941017}. One high energy (18\,GeV) photon was observed in GRB\,940217 90 minutes after the burst trigger \citep{hurl}. The RHESSI burst GRB\,021206 also seems to have an excess at high energies \citep[e.g.][]{wigger07}. \citet{kaneko08} analysed 15 BATSE GRBs with possible high energy components observed by TASC, the Total Absorption Shower Counter on EGRET, including GRB\,941017. They found that high energy components were present for two bursts in the sample, and a third burst had a probable peak energy in excess of 170\,MeV. High energy photons between 25--50\,MeV have also been observed in the AGILE burst GRB\,080514b \citep{agile08}. These photons occurred after the apparent end of the hard X-ray emission. This high energy emission has so far been observed solely in a few bright bursts, indicating that it may be relatively unusual among GRBs.

A possible interpretation of the high energy component is that it is due to synchrotron self-Compton (SSC) emission from the reverse shock \citep{gg03}, where the synchrotron emitting electrons are responsible for the low energy spectrum. An inverse Compton peak can be observed at 10\,MeV--100\,GeV, which is delayed relative to the softer emission and has a longer decay time \citep{sp04}. However, the emission is not delayed in GRB\,061122, but is present throughout the burst and for a small interval after the burst, which may rule out SSC emission. A possible emission mechanism to provide the right temporal behaviour is the reverse shock which travels into the GRB ejecta as it is decelerated by the circumburst medium. Internal shocks seem to be ruled out, since the synchrotron self-Compton spectrum implies $\epsilon_B \sim 10^{-7}$ (the fraction of internal energy behind the shock in the magnetic field). This value is much lower than that expected from the magnetic field advected by the ejecta from the source. 

Recently, a high energy component has been proposed in the spectrum of GRB\,080319B as an explanation for the prompt optical and $\gamma$--ray emission \citep{racusin}. The optical emission during the prompt phase was up to $10^4$ times greater than the extrapolated $\gamma$--ray flux, leading to the conclusion that synchrotron radiation was responsible for the optical emission and SSC radiation was responsible for the soft $\gamma$--ray spectrum. It was proposed that a third spectral component was present at GeV energies, due to second order Compton scattering, and that most of the energy of the burst was emitted at high energies.

\citet{dermer2000} performed calculations of prompt and afterglow GRB emission using the standard blast-wave model with $\Gamma_0 \sim 300$. They proposed that the high energy emission from GRB\,940217 during the burst and at late times were the result of non-thermal synchrotron and self-synchroton Compton (SSC) emission moving through the GeV band respectively. Calculations were also performed for a ``clean" fireball ($\Gamma_0 \sim 1000$) and a ``dirty" fireball ($\Gamma_0 \sim 100$) to investigate the peak flux emission from MeV--TeV energies. The clean fireball model predicts brief MeV emission at the start of the burst, with the burst having a large peak flux and high $E_{peak}$, while the dirty fireball predicts later MeV emission and a weaker burst with a low $E_{peak}$. The standard model predicted the peak in MeV emission at $t \sim 4$\,s and a luminosity at MeV energies of $\sim 1.5 \times 10^{51}$ erg\,s$^{-1}$ at $pz = 0.95$ \citep{dermer2000}. GRB\,061122 is consistent with the standard model, with a luminosity from 1--8\,MeV of $\sim 7 \times 10^{50}$ erg\,s$^{-1}$.  The standard model seems to be favoured over the alternatives because  its duration is more extended than the dirty fireball and background effects are more important in the clean fireball model. 

\citet{rr04} has argued that a continually decreasing post-burst relativistic outflow may exist for some GRBs, caused by the sluggish infall of matter into a compact object. It can be reprocessed by the soft photon field radiation and produce high energy $\gamma$--rays, thus providing energy injection on a much larger timescale than the apparent duration of the burst. The Compton Drag process mentioned above could be very effective in extracting energy from the relativistic wind.

\subsection{Afterglow Properties\label{ag2}}
Monochromatic breaks in the afterglow lightcurve can be used to estimate the opening angle of the jet producing the emission. These breaks are observed when the Lorentz factor $\Gamma$ drops below the inverse of the jet angle $\theta_{j}$ so that the radiation is beamed outside the original jet, reducing the observed flux \citep{rhoads99, piran04}. No break was observed in the X--ray lightcurve of GRB\,061122. Observations did not start until $\sim$ 7 hours after the trigger, so it can be assumed that the jet break occurred before the onset of the XRT observation. Setting an upper limit on the jet break time of 24.5\,ks (Sect.~\ref{ag}), this implies a limit on the jet opening angle of 2.8$^{\circ}$, using equation (1) from \citet{frail2001}, assuming an ISM density of 1 cm$^{-3}$ and a pseudo redshift of 0.95. Similarly, if the jet break occurred after the end of the observation (76\,ks), an limit of 11.9$^{\circ}$ can be derived for the jet opening angle. Therefore the jet angle must be either smaller than 2.8$^{\circ}$ or larger than 11.9$^{\circ}$.

GRB\,061122 had an optical afterglow with $R_{mag} \sim 23$, consistent with the apparent magnitudes measured for a large sample of long GRBs detected by Swift and other missions at 1 day and 4 days after the burst, corrected to a common z=1 system \citep{kann07,kann08}. 

\subsection{Constraints on Polarisation\label{disc_pol}}
Two possible explanations for a significant level of polarisation are synchrotron radiation and Compton Drag. Synchrotron radiation, from an ordered magnetic field advected from the central engine \citep{lyut03}, is a general feature of GRBs. The level of polarisation produced by a perfectly ordered magnetic field can be $\Pi_{s} = (p + 1)/(p + 7/3)$ where \emph{p} represents the electron distribution power--law index. Typical values of \emph{p} = 2--3 correspond to a percentage polarisation of 70--75\%. However, this high level is not observed in GRB\,061122. Compton Drag, which occurs when photons are inverse Compton scattered and are beamed in an opening angle $\sim 1/\Gamma$ \citep{lazz04}, can also produce a significant level of
polarisation. An alternative scenario for polarisation occurs when a jet with a small opening angle is viewed
slightly off--axis \citep{waxman03}. 

\citet{lazz04} calculated the polarisation via Compton Drag as a function of the
observer angle for several jet geometries, and showed that
polarisation can be produced if the condition $\Gamma\theta_{j} \leq 5$
is satisfied, where $\Gamma$ is the Lorentz factor of the jet and $\theta_{j}$ is the opening angle of the jet. GRB\,061122 has an isotropic energy of $8.5\,\times$\,10$^{52}$ erg (\S\,\ref{rl}). The Lorentz factor of
the fireball can be obtained from the redshift corrected peak energy of GRB\,061122 (E$_{peak,z} = 330$~keV) by the relationship

\begin{equation}
E_{peak} \simeq 10~\Gamma^{2}\,kT
\label{eqn:epeak}
\end{equation}

where T$\sim$10$^{5}$~K is the black body spectrum of the photon field
\citep{lazz04}. The computed value for GRB\,061122 is $\Gamma \sim$ 62. This value is relatively low, compared to previous measurements in the range $\Gamma \sim 100-400$ for bursts with high energy photons \cite[e.g.][]{ls2001}. Using the estimated values for the jet opening angles derived in \S\,\ref{ag2} ($\theta_{j} < 2.8^{\circ}$ or $>11.9^{\circ}$) yield the respective results:

\begin{equation}
\Gamma\theta_{j} \leq 3
\label{eqn:result1}
\end{equation}
or

\begin{equation}
\Gamma\theta_{j} \leq 13.
\label{eqn:result1}
\end{equation}

The smaller opening angle fulfills the condition for polarisation, whereas the larger angle does not. Therefore, an upper limit on the polarisation is the firmest conclusion that we can draw from the data.

In the fireball model, the fractional polarisation emitted by each
  element remains the same, but the direction of the polarisation vector of the
  radiation emitted by different elements within the shell is rotated by
  different amounts. This can lead to effective depolarisation of the total
  emission \citep{lyut03}.

\section{Conclusions\label{conc}}
GRB\,061122 is one of the brightest gamma-ray bursts observed by \textit{INTEGRAL} to date, with a fluence (20--200\,keV) of $\sim 10^{-5}$ erg\,cm$^{-2}$. The afterglow of GRB\,061122 was observed by the XRT on \textit{Swift} and optical observations were also carried out. The pseudo--redshift calculated for GRB\,061122 is $pz = 0.95 \pm 0.18$. The values of L$_{iso}$ and E$_{iso}$ were determined for GRB\,061122 resulting in L$_{iso}$ = $1.47 \pm 0.05 \times 10^{52}$ ergs s$^{-1}$ and E$_{iso}$ = $8.5 \pm 3.6\,\times$ 10$^{52}$ erg. 

An upper polarisation limit of 60\% was determined for GRB\,061122. A more definite value could not be obtained due to lack of statistics. Assuming that the jet break occurred outside the observation time of XRT, the jet opening angle must be either smaller than 2.8$^{\circ}$or larger than 11.9$^{\circ}$. Using these limits, the conditions for polarisation could be fulfilled if $\theta_{j} \lesssim 2.8^{\circ}$. 

GRB\,061122 exhibited a high energy spectral component in the observed $\gamma$--ray spectrum. The high energy component does not turn over within the energy range of SPI, indicating that emission may exist above $\sim$\,8\,MeV. GRB\,061122 seems most consistent with the standard blast-wave model as proposed by \citet{dermer2000}, with a luminosity from 1--8\,MeV of $\sim 7 \times 10^{50}$ erg\,s$^{-1}$. High energy missions such as the Fermi Gamma-ray Space Telescope \citep{glast}, launched in June 2008, have a wider energy range (up to $\sim$\,300\,GeV). Therefore, Fermi will provide a better picture of the occurrence of high energy components in GRB spectra and differentiate between different spectral models.

\begin{acknowledgements}
S.McG. would like to acknowledge support from the Swedish National Space Board. S.M.B. acknowledges European Union support through a Marie Curie Fellowship in FP6.
\end{acknowledgements}

\bibliography{refs}

\hyphenation{Post-Script Sprin-ger}
\begin{thebibliography}{53}
\expandafter\ifx\csname natexlab\endcsname\relax\def\natexlab#1{#1}\fi

\bibitem[{Agostinelli {et~al.}(2003)Agostinelli, Allison, Amako,
  {et~al.}}]{Agostinelli}
Agostinelli, S., Allison, J., Amako, K., {et~al.} 2003, Nucl. Instrum. Methods
  Phys. Res., Sect. A, 506, 250

\bibitem[{{Band} {et~al.}(1993){Band}, {Matteson}, {Ford}, {Schaefer},
  {Palmer}, {Teegarden}, {Cline}, {Briggs}, {Paciesas}, {Pendleton}, {Fishman},
  {Kouveliotou}, {Meegan}, {Wilson}, \& {Lestrade}}]{band:1993}
{Band}, D., {Matteson}, J., {Ford}, L., {et~al.} 1993, \apj, 413, 281

\bibitem[{{Daigne} \& {Mochkovitch}(2002)}]{dm02}
{Daigne}, F. \& {Mochkovitch}, R. 2002, \mnras, 336, 1271

\bibitem[{{de Angelis}(2001)}]{glast}
{de Angelis}, A. 2001, in New Worlds in Astroparticle Physics, ed. A.~M.
  {Mourao}, M.~{Pimenta}, P.~M. {Sa}, \& J.~M. {Velhinho}, 140

\bibitem[{{Dermer} {et~al.}(2000){Dermer}, {Chiang}, \& {Mitman}}]{dermer2000}
{Dermer}, C.~D., {Chiang}, J., \& {Mitman}, K.~E. 2000, \apj, 537, 785

\bibitem[{{Diehl} {et~al.}(2003){Diehl}, {Baby}, {Beckmann}, {Connell},
  {Dubath}, {Jean}, {Kn{\"o}dlseder}, {Roques}, {Schanne}, {Shrader},
  {Skinner}, {Strong}, {Sturner}, {Teegarden}, {von Kienlin}, \&
  {Weidenspointner}}]{diehl2003}
{Diehl}, R., {Baby}, N., {Beckmann}, V., {et~al.} 2003, \aap, 411, L117

\bibitem[{{Ferguson} {et~al.}(2003){Ferguson}, {Barlow}, {Bird}, {Dean},
  {Hill}, {Shaw}, {Stephen}, {Sturner}, {Tikkanen}, {Weidenspointner}, \&
  {Willis}}]{2003A&A...411L..19F}
{Ferguson}, C., {Barlow}, E.~J., {Bird}, A.~J., {et~al.} 2003, \aap, 411, L19

\bibitem[{{Foley} {et~al.}(2008){Foley}, {McGlynn}, {Hanlon}, {McBreen}, \&
  {McBreen}}]{sf07}
{Foley}, S., {McGlynn}, S., {Hanlon}, L., {McBreen}, S., \& {McBreen}, B. 2008,
  \aap, 484, 143

\bibitem[{{Frail} {et~al.}(2001){Frail}, {Kulkarni}, {Sari}, {Djorgovski},
  {Bloom}, {Galama}, {Reichart}, {Berger}, {Harrison}, {Price}, {Yost},
  {Diercks}, {Goodrich}, \& {Chaffee}}]{frail2001}
{Frail}, D.~A., {Kulkarni}, S.~R., {Sari}, R., {et~al.} 2001, \apjl, 562, L55

\bibitem[{{Ghirlanda} {et~al.}(2003){Ghirlanda}, {Celotti}, \&
  {Ghisellini}}]{ghirlanda:2003}
{Ghirlanda}, G., {Celotti}, A., \& {Ghisellini}, G. 2003, in American Institute
  of Physics Conference Series, Vol. 662, Gamma-Ray Burst and Afterglow
  Astronomy 2001: A Workshop Celebrating the First Year of the HETE Mission,
  ed. G.~R. {Ricker} \& R.~K. {Vanderspek}, 270--272

\bibitem[{{Giuliani} {et~al.}(2008){Giuliani}, {Mereghetti}, {Fornari}, {Del
  Monte}, {Feroci}, {Marisaldi}, {Esposito}, {Perotti}, {Tavani}, {Argan},
  {Barbiellini}, {Boffelli}, {Bulgarelli}, {Caraveo}, {Cattaneo}, {Chen},
  {Costa}, {D'Ammando}, {di Cocco}, {Donnarumma}, {Evangelista}, {Fiorini},
  {Fuschino}, {Galli}, {Gianotti}, {Labanti}, {Lapshov}, {Lazzarotto},
  {Lipari}, {Longo}, {Morselli}, {Pacciani}, {Pellizzoni}, {Piano}, {Picozza},
  {Prest}, {Pucella}, {Rapisarda}, {Rappoldi}, {Soffitta}, {Trifoglio},
  {Trois}, {Vallazza}, {Vercellone}, {Zanello}, {Salotti}, {Cutini}, {Pittori},
  {Preger}, {Santolamazza}, {Verrecchia}, {Gehrels}, {Page}, {Burrows},
  {Rossi}, {Hurley}, {Mitrofanov}, \& {Boynton}}]{agile08}
{Giuliani}, A., {Mereghetti}, S., {Fornari}, F., {et~al.} 2008, \aap, 491, L25

\bibitem[{{Golenetskii} {et~al.}(2006){Golenetskii}, {Aptekar}, {Mazets},
  {Pal'shin}, {Frederiks}, \& {Cline}}]{gcn5841}
{Golenetskii}, S., {Aptekar}, R., {Mazets}, E., {et~al.} 2006, {GCN 5841}

\bibitem[{{Golenetskii} {et~al.}(1998){Golenetskii}, {Aptekar}, {Frederiks},
  {Il'Inskii}, {Mazets}, {Panov}, {Sokolova}, \& {Terekhov}}]{konus}
{Golenetskii}, S.~V., {Aptekar}, R.~L., {Frederiks}, D.~D., {et~al.} 1998, in
  American Institute of Physics Conference Series, Vol. 428, American Institute
  of Physics Conference Series, ed. C.~A. {Meegan}, R.~D. {Preece}, \& T.~M.
  {Koshut}, 284

\bibitem[{{Gonz{\'a}lez} {et~al.}(2003){Gonz{\'a}lez}, {Dingus}, {Kaneko},
  {Preece}, {Dermer}, \& {Briggs}}]{941017}
{Gonz{\'a}lez}, M.~M., {Dingus}, B.~L., {Kaneko}, Y., {et~al.} 2003, \nat, 424,
  749

\bibitem[{{Granot} \& {Guetta}(2003)}]{gg03}
{Granot}, J. \& {Guetta}, D. 2003, \apjl, 598, L11

\bibitem[{{Halpern}(2006)}]{gcn5849}
{Halpern}, J. 2006, {GCN 5849}

\bibitem[{{Hurley} {et~al.}(1994){Hurley}, {Dingus}, {Mukherjee}, {Sreekumar},
  {Kouveliotou}, {Meegan}, {Fishman}, {Band}, {Ford}, {Bertsch}, {Cline},
  {Fichtel}, {Hartman}, {Hunter}, {Thompson}, {Kanbach}, {Mayer-Hasselwander},
  {von Montigny}, {Sommer}, {Lin}, {Nolan}, {Michelson}, {Kniffen}, {Mattox},
  {Schneid}, {Boer}, \& {Niel}}]{hurl}
{Hurley}, K., {Dingus}, B.~L., {Mukherjee}, R., {et~al.} 1994, \nat, 372, 652

\bibitem[{{Kalemci} {et~al.}(2007){Kalemci}, {Boggs}, {Kouveliotou},
  {et~al.}}]{kal07}
{Kalemci}, E., {Boggs}, S.~E., {Kouveliotou}, C., {et~al.} 2007, \apjs, 169, 75

\bibitem[{{Kaneko} {et~al.}(2008){Kaneko}, {Gonz{\'a}lez}, {Preece}, {Dingus},
  \& {Briggs}}]{kaneko08}
{Kaneko}, Y., {Gonz{\'a}lez}, M.~M., {Preece}, R.~D., {Dingus}, B.~L., \&
  {Briggs}, M.~S. 2008, \apj, 677, 1168

\bibitem[{{Kaneko} {et~al.}(2006){Kaneko}, {Preece}, {Briggs}, {Paciesas},
  {Meegan}, \& {Band}}]{yuki06}
{Kaneko}, Y., {Preece}, R.~D., {Briggs}, M.~S., {et~al.} 2006, \apjs, 166, 298

\bibitem[{{Kann} {et~al.}(2007){Kann}, {Klose}, {Zhang}, {Malesani}, {Nakar},
  {Wilson}, {Butler}, {Antonelli}, {Chincarini}, {Cobb}, {Covino}, {D'Avanzo},
  {D'Elia}, {Della Valle}, {Ferrero}, {Fugazza}, {Gorosabel}, {Israel},
  {Mannucci}, {Piranomonte}, {Schulze}, {Stella}, {Tagliaferri}, \&
  {Wiersema}}]{kann07}
{Kann}, D.~A., {Klose}, S., {Zhang}, B., {et~al.} 2007, astro-ph/0712.2186

\bibitem[{{Kann} {et~al.}(2008){Kann}, {Klose}, {Zhang}, {Wilson}, {Butler},
  {Malesani}, {Nakar}, {Antonelli}, {Chincarini}, {Cobb}, {Covino}, {D'Avanzo},
  {D'Elia}, {Della Valle}, {Ferrero}, {Fugazza}, {Gorosabel}, {Israel},
  {Mannucci}, {Piranomonte}, {Schulze}, {Stella}, {Tagliaferri}, \&
  {Wiersema}}]{kann08}
{Kann}, D.~A., {Klose}, S., {Zhang}, B., {et~al.} 2008, astro-ph/0804.1959

\bibitem[{{Lazzati}(2006)}]{lazz2006}
{Lazzati}, D. 2006, New Journal of Physics, 8, 131

\bibitem[{Lazzati {et~al.}(2004)Lazzati, Rossi, Ghisellini, \& Rees}]{lazz04}
Lazzati, D., Rossi, E., Ghisellini, G., \& Rees, M.~J. 2004, \mnras, 347, L1

\bibitem[{{Lithwick} \& {Sari}(2001)}]{ls2001}
{Lithwick}, Y. \& {Sari}, R. 2001, \apj, 555, 540

\bibitem[{{Lyutikov} {et~al.}(2003){Lyutikov}, {Pariev}, \&
  {Blandford}}]{lyut03}
{Lyutikov}, M., {Pariev}, V.~I., \& {Blandford}, R.~D. 2003, \apj, 597, 998

\bibitem[{{Lyutikov} \& {Usov}(2000)}]{lyut00}
{Lyutikov}, M. \& {Usov}, V.~V. 2000, \apjl, 543, L129

\bibitem[{{McBreen} {et~al.}(2006{\natexlab{a}}){McBreen}, {Beardmore},
  {Oates}, {Page}, {Barthelmy}, {Burrows}, {Roming}, \& {Gehrels}}]{rep_17}
{McBreen}, S., {Beardmore}, A.~P., {Oates}, S.~R., {et~al.} 2006{\natexlab{a}},
  GCN Reports, 17

\bibitem[{{McBreen} {et~al.}(2006{\natexlab{b}}){McBreen}, {Hanlon}, {McGlynn},
  {McBreen}, {Foley}, {Preece}, {von Kienlin}, \& {Williams}}]{mcbreen06}
{McBreen}, S., {Hanlon}, L., {McGlynn}, S., {et~al.} 2006{\natexlab{b}}, \aap,
  455, 433

\bibitem[{{McGlynn} {et~al.}(2007){McGlynn}, {Clark}, {Dean}, {Hanlon},
  {McBreen}, {Willis}, {McBreen}, {Bird}, \& {Foley}}]{me06}
{McGlynn}, S., {Clark}, D.~J., {Dean}, A.~J., {et~al.} 2007, \aap, 466, 895

\bibitem[{{McGlynn} {et~al.}(2008){McGlynn}, {Foley}, {McBreen}, {Hanlon},
  {O'Connor}, {Carrillo}, \& {McBreen}}]{me08}
{McGlynn}, S., {Foley}, S., {McBreen}, S., {et~al.} 2008, \aap, 486, 405

\bibitem[{{Mereghetti} \& {G{\"o}tz}(2006)}]{gcn5836}
{Mereghetti}, S. \& {G{\"o}tz}, D. 2006, {GCN 5836}

\bibitem[{Mereghetti {et~al.}(2003)Mereghetti, {G\"{o}tz}, Borkowski,
  {et~al.}}]{ibas}
Mereghetti, S., {G\"{o}tz}, D., Borkowski, J., {et~al.} 2003, \aap, 411, L291

\bibitem[{{Mereghetti} {et~al.}(2006){Mereghetti}, {Paizis}, {G{\"o}tz},
  {Petry}, {Mowlavi}, {Beck}, \& {Borkowski}}]{gcn5834}
{Mereghetti}, S., {Paizis}, A., {G{\"o}tz}, D., {et~al.} 2006, {GCN 5834}

\bibitem[{M{\'e}sz{\'a}ros(2006)}]{mesz2006}
M{\'e}sz{\'a}ros, P. 2006, Reports of Progress in Physics, 69, 2259

\bibitem[{{Oates} \& {McBreen}(2006)}]{gcn5846}
{Oates}, S.~R. \& {McBreen}, S. 2006, {GCN 5846}

\bibitem[{{P{\'e}langeon} {et~al.}(2008){P{\'e}langeon}, {Atteia}, {Nakagawa},
  {Hurley}, {Yoshida}, {Vanderspek}, {Suzuki}, {Kawai}, {Pizzichini},
  {Bo{\"e}r}, {Braga}, {Crew}, {Donaghy}, {Dezalay}, {Doty}, {Fenimore},
  {Galassi}, {Graziani}, {Jernigan}, {Lamb}, {Levine}, {Manchanda}, {Martel},
  {Matsuoka}, {Olive}, {Prigozhin}, {Ricker}, {Sakamoto}, {Shirasaki},
  {Sugita}, {Takagishi}, {Tamagawa}, {Villasenor}, {Woosley}, \&
  {Yamauchi}}]{pel08}
{P{\'e}langeon}, A., {Atteia}, J.-L., {Nakagawa}, Y.~E., {et~al.} 2008, \aap,
  491, 157

\bibitem[{Piran(2004)}]{piran04}
Piran, T. 2004, Reviews of Modern Physics, 76, 1143

\bibitem[{{Racusin} {et~al.}(2008){Racusin}, {Karpov}, {Sokolowski}, {Granot},
  {Wu}, {Pal'Shin}, {Covino}, {van der Horst}, {Oates}, {Schady}, {Smith},
  {Cummings}, {Starling}, {Piotrowski}, {Zhang}, {Evans}, {Holland}, {Malek},
  {Page}, {Vetere}, {Margutti}, {Guidorzi}, {Kamble}, {Curran}, {Beardmore},
  {Kouveliotou}, {Mankiewicz}, {Melandri}, {O'Brien}, {Page}, {Piran},
  {Tanvir}, {Wrochna}, {Aptekar}, {Barthelmy}, {Bartolini}, {Beskin}, {Bondar},
  {Bremer}, {Campana}, {Castro-Tirado}, {Cucchiara}, {Cwiok}, {D'Avanzo},
  {D'Elia}, {Valle}, {de Ugarte Postigo}, {Dominik}, {Falcone}, {Fiore}, {Fox},
  {Frederiks}, {Fruchter}, {Fugazza}, {Garrett}, {Gehrels}, {Golenetskii},
  {Gomboc}, {Gorosabel}, {Greco}, {Guarnieri}, {Immler}, {Jelinek},
  {Kasprowicz}, {La Parola}, {Levan}, {Mangano}, {Mazets}, {Molinari},
  {Moretti}, {Nawrocki}, {Oleynik}, {Osborne}, {Pagani}, {Pandey}, {Paragi},
  {Perri}, {Piccioni}, {Ramirez-Ruiz}, {Roming}, {Steele}, {Strom}, {Testa},
  {Tosti}, {Ulanov}, {Wiersema}, {Wijers}, {Winters}, {Zarnecki}, {Zerbi},
  {M{\'e}sz{\'a}ros}, {Chincarini}, \& {Burrows}}]{racusin}
{Racusin}, J.~L., {Karpov}, S.~V., {Sokolowski}, M., {et~al.} 2008, \nat, 455,
  183

\bibitem[{{Ramirez-Ruiz}(2004)}]{rr04}
{Ramirez-Ruiz}, E. 2004, \mnras, 349, L38

\bibitem[{{Rhoads}(1999)}]{rhoads99}
{Rhoads}, J.~E. 1999, \apj, 525, 737

\bibitem[{{Ryde}(2004)}]{ryde:2004}
{Ryde}, F. 2004, \apj, 614, 827

\bibitem[{{Ryde}(2005)}]{ryde05}
{Ryde}, F. 2005, \apjl, 625, L95

\bibitem[{{Skinner} \& {Connell}(2003)}]{skin2003}
{Skinner}, G. \& {Connell}, P. 2003, \aap, 411, L123

\bibitem[{{Stern} \& {Poutanen}(2004)}]{sp04}
{Stern}, B.~E. \& {Poutanen}, J. 2004, \mnras, 352, L35

\bibitem[{Ubertini {et~al.}(2003)Ubertini, Lebrun, Cocco, {et~al.}}]{uber2003}
Ubertini, P., Lebrun, F., Cocco, G.~D., {et~al.} 2003, \aap, 411, L131

\bibitem[{Vedrenne {et~al.}(2003)Vedrenne, Roques, {Sch\"{o}nfelder},
  {et~al.}}]{ved2003}
Vedrenne, G., Roques, J.~P., {Sch\"{o}nfelder}, V., {et~al.} 2003, \aap, 411,
  L63

\bibitem[{{Vianello} {et~al.}(2008){Vianello}, {G{\"o}tz}, \&
  {Mereghetti}}]{vian08}
{Vianello}, G., {G{\"o}tz}, D., \& {Mereghetti}, S. 2008, astro-ph/0812.3349

\bibitem[{{Waxman}(2003)}]{waxman03}
{Waxman}, E. 2003, \nat, 423, 388

\bibitem[{{Wigger} {et~al.}(2004){Wigger}, {Hajdas}, {Arzner}, {G{\"u}del}, \&
  {Zehnder}}]{wigger04}
{Wigger}, C., {Hajdas}, W., {Arzner}, K., {G{\"u}del}, M., \& {Zehnder}, A.
  2004, \apj, 613, 1088

\bibitem[{{Wigger} {et~al.}(2008){Wigger}, {Wigger}, {Bellm}, \&
  {Hajdas}}]{wigger07}
{Wigger}, C., {Wigger}, O., {Bellm}, E., \& {Hajdas}, W. 2008, \apj, 675, 553

\bibitem[{Winkler {et~al.}(2003)Winkler, {Courvoisier}, {Di Cocco},
  {et~al.}}]{wink2003}
Winkler, C., {Courvoisier}, T.~J.~L., {Di Cocco}, G., {et~al.} 2003, \aap, 411,
  L1

\bibitem[{{Zhang} \& {M{\'e}sz{\'a}ros}(2004)}]{Zhang2004}
{Zhang}, B. \& {M{\'e}sz{\'a}ros}, P. 2004, International Journal of Modern
  Physics A, 19, 2385

\end{thebibliography}
\bibliographystyle{aa}
\end{document}